# Thermoelectric and lattice dynamics properties of layered MX (M = Sn, Pb; X = S, Te) compounds

Abhiyan Pandit*,[a] and Bothina Hamad[a,b]

[a]*Physics Department, University of Arkansas, Fayetteville, AR 72701, USA*
[b]*Physics Department, The University of Jordan, Amman-11942, Jordan*

**Abstract**

Lead and tin chalcogenides have been studied widely due to their promising thermoelectric (TE) properties. Further enhancement in their TE efficiency has been reported upon the reduction of the dimension, which is an important feature in modern device fabrications. Using density functional theory combined with the Semi-classical Boltzmann transport theory, we studied the structural, electronic and TE properties of two-dimensional (2D) MX (M = Sn, Pb; X = S, Te) monolayers (MLs). Spin-orbit coupling was found to have significant effects on their electronic structure, particularly for the heavy compounds. Structural optimization followed by phonon transport studies prevailed that the rectangular (γ-) phase is energetically the most favorable for SnS and SnTe MLs, whereas the square structure is found the most stable for PbS and PbTe MLs. Our results are in a good agreement with previous studies. These 2D materials exhibit high Seebeck coefficients (1000 - 1500 µV/K) and power factors ((33 - 77.3) × $10^{-4}$ $Wm^{-1}K^{-2}$) along with low lattice thermal conductivities (< 3 $Wm^{-1}K^{-1}$), which are essential features of good TE materials. The maximum figure of merits (*ZT*) of 1.04, 1.46, 1.51 and 1.94 are predicted for *n*-type SnS, SnTe, PbS and *p*-type PbTe MLs, respectively at 700 K, which are higher than their bulk *ZT* values. Hence, these MLs are promising candidates for TE applications.

**Keywords:** *Two-dimensional monochalcogenides, phonon dispersion, lattice thermal conductivity, thermoelectric properties.*

I. Introduction

The demand for energy is increasing widely due to the surge in the economic growth, which increases the risk of future scarcity of natural energy resources and global warming [1,2]. This fact has urged the academic governmental authorities and industries to seek for new resources that has to be renewable, sustainable and environment friendly. One of the promising renewable resources is the thermoelectric (TE) energy. TE devices are composed of two dissimilar semiconductors (namely, *p*- and *n*-type) that are connected electrically in series and thermally in parallel. These devices can operate for a direct conversion of thermal energy into electricity or vice versa, which are driven by three inter-related phenomena; Seebeck, Peltier and the Thomson effects [3–5]. The TE performance of a material is measured by a dimensionless quantity *ZT*, called the figure of merit:

$$ZT = \frac{S^2 \sigma T}{k}, \qquad (1)$$

where $S$ is the Seebeck coefficient, $\sigma$ is the electrical conductivity, $T$ is temperature, and $\kappa$ is the thermal conductivity (sum of the electronic thermal conductivity ($\kappa_e$) and the lattice thermal conductivity ($\kappa_l$)). It is important to note that improving the value of *ZT* has been a challenging task because of the inter-correlated individual parameters contributing to *ZT*, which in turn give rise to a low energy conversion efficiency. High quality PbTe samples, which is the most studied compound in the group IV-VI, with optimal doping is found to have inherently high TE efficiency and further increases with the band structure engineering [5–10]. The least studied compound in the group IV-VI is PbS, which was also found to have a significantly high *ZT* ~ 0.8 in its inherent form that increases to ~ 1 upon delicate manipulation of the nanostructured secondary phase [11–14]. In addition, *p*-type doped SnTe with Ag has shown a high TE efficiency of ~ 1.23 [15]. Other monochalcogenides such as SnS, GeSe, GeS and GeTe were also theoretically predicted to have a good TE performance [16,17].

Since the discovery of graphene, two-dimensional (2D) materials have been extensively studied regarding their novel structural, electronic, optical and TE properties due to their potential photovoltaic, optoelectronic and transistor based applications [18–26]. Typical 2D materials such as group IV-VI chalcogenides are found to have relatively high electrical properties such as the Seebeck coefficient and electrical conductivity. They also possess a low lattice thermal conductivity, which leads to a reasonably good TE efficiency. Nano-structuring of the bulk materials has been investigated to be efficient in reducing the lattice thermal conductivities that leads to higher TE figure of merit ($ZT$) [11,27,28]. In addition, it was found that the Seebeck coefficient increases as a result of higher density of states near Fermi level in low dimensional materials, which corresponds to the higher efficiency [29–33]. Recent reports on 2D monolayers (MLs) of SnS and SnSe showed optimal $ZT$ values of 1.88 and 2.63 - 3.27, respectively [34,35]. Another study predicted higher ZT values over 2.5 for SnX (X=Te, Se, S) MLs in their hexagonal (β-) phase [36]. These all facts make them promising candidates for TE applications.

This work is devoted to study the effect on the TE properties of lead (Pb) and tin (Sn) chalcogenides upon the reduction of dimensionality from their bulk to the most stable 2D phases. In the recent studies, tin chalcogenide (SnS and SnTe) MLs were found to crystallize in γ-(rectangular) phase, whereas lead chalcogenide (PbS and PbTe) MLs were found to crystallize in the square structure at room temperature [37,38]. In specific, γ-phase of SnTe was observed to be the most stable structure at the mono- and bi-layered limit [39,40]. Similar to previous studies [35,41,42], a temperature range from 300 K to 700 K has been adopted. A significantly high TE efficiency is predicted in these group IV-VI compounds as they are found to have low lattice thermal conductivity. The rest of the paper is organized as: Section II includes the computational method; Section III presents the results and discussion; and Section IV summarizes the concluding remarks.

## II. Computational Methods

Structural and electronic calculations are performed using density functional theory (DFT) based on first principles as implemented in the VASP by including the van der Waals interaction [43–47]. The generalized gradient approximation (GGA) using the projector augmented wave method with Perdew-Burke-Ernzerhof (PAW-PBE) functional was used for the exchange and correlation effects [48]. A cut off energy of 450 eV was used with the atomic plane and its neighboring images separated by a 15 Å vacuum in the z-direction. The convergence criterion for energy was set to $10^{-6}$ eV. Primitive unit cells with $k$-meshes of 18 × 18 × 1 and 36 × 36 × 1 were used to compute the structural relaxation and density of sates/transport calculations, respectively.

The transport coefficients (S, $\sigma$ and $\kappa_e$) are obtained by using semi-classical Boltzmann theory as implemented in the BoltzTraP2 code (version-19.7.3) [49], which uses the inputs from DFT calculations based on the rigid band approximation (RBA). The Seebeck coefficient (S) is independent of the relaxation time ($\tau$), whereas $\sigma$ and $\kappa_e$ are calculated relative to $\tau$. Under the constant relaxation time approximation, the relaxation time ($\tau$) is set to $1.0 \times 10^{-14}$ s which was the value adopted in the similar calculations [16,42,50]. The lattice thermal conductivity ($\kappa_l$) was calculated using the second-order (harmonic) and third-order (anharmonic) interatomic force constants (IFCs) as implemented in the package ShengBTE [51]. To calculate the second-order IFCs, PHONOPY code [52] was used with a supercell of 5 × 5 × 1 and $k$-mesh of a 5 × 5 × 1. The force constants ($C_{i\alpha;j\beta}$) are then obtained from the forces that result from the displacements in a system as follows:

$$C_{i\alpha;j\beta} = -\frac{F_{i\alpha}}{\Delta_{j\beta}}, \qquad (2)$$

where $F_{i\alpha}$ is the force acting on atom $i$ along the $\alpha$ direction due to the displacement $\Delta_{j\beta}$ (along the $\beta$ direction of atom $j$). The displacement amplitude of 0.01 Å was applied for the atom along $\pm x$, $\pm y$ and $\pm z$ directions. A nearest neighbor interaction up to the 7$^{th}$ nearest neighbors with a supercell of 5 × 5 × 1 was set up for the calculation of third-order IFCs.

### III. Results and Discussions

#### A. Structural and Electronic Properties

This subsection presents the structural and electronic properties of the 2D MX (M = Sn, Pb; X = S, Te) MLs along with their bulk analogs. The structural optimizations showed that all of them in their bulk form crystallize in the cubic structure (space group of $Fm\bar{3}m$) except SnS, which has an orthorhombic structure (space group of *Pnma*). The MLs of PbTe (PbS) and SnS retain the square and orthorhombic (rectangular) structures, respectively as shown in Fig. 1. However, SnTe is transformed to rectangular in its ML limit. The calculated lattice parameters are summarized in Table I, which are in good agreement with previous results.

The electronic band structure and total density of states (DOS) of SnS, PbS, SnTe and PbTe MLs including spin-orbit coupling (SOC) are shown in Fig. 2 (their bulk electronic structure in Fig. S1). It is found that SnS has indirect band gap, while SnTe, PbS and PbTe MLs have direct band gaps. The values of electronic band gap for each of the layered compounds (with and without SOC) and their bulk forms (with SOC) are listed in Table II along with previous reports. These MLs have larger band gaps than their corresponding bulk forms, which is partially due to the quantum confinement effect [60]. These changes in the band gaps will affect the TE properties.

The band gaps (without SOC) of PbS and PbTe MLs are close to previous results. However, SOC is an intrinsic property of atoms/compounds and known to have a crucial effect that breaks the degeneracy of band dispersion and thereby affects the band gap [61,62]. This is clearly noticed for the investigated compounds that have heavy elements (PbTe, PbS and SnTe). Hence, the SOC was considered for TE calculations.

**B. Lattice Dynamics Properties**

Lattice dynamics properties are studied in order to investigate the structural stability of the discussed materials and to evaluate their lattice thermal conductivity ($\kappa_l$) values. The phonon band structure of these MX (M = Sn, Pb; X= S, Te) MLs are shown in Fig. 3(a, d, g, j). The positive frequencies of these compounds indicate that they are dynamically stable. As the unit cell has 4 atoms, there are total 12 (3N) modes of vibrations, where the lowest three branches correspond to the acoustic modes and rest nine are the optical modes. The phonon DOS of these MLs show that Pb/Sn/Te atoms have relatively lower frequencies than S atoms (see Fig. 3(b, e, h, k)). This can be attributed to the heavier masses of Pb/Sn/Te atoms than that of S atom. The group velocity is defined as the slope of the dispersion relation, given as $v_g = \partial\omega(k)/\partial k$, which is an important factor in in determining the $\kappa_l$ as $\kappa_l \propto v_g$ (see Eq. 3). From the Fig. 3(c, f, i, l), it is evident that the intermediate optical branches show a significant contribution to the group velocity, in agreement with previous studies [41,59,65,66]. In addition to $v_g$, the specific heat capacity ($C_v$) is also an important thermal property of a material required to estimate $\kappa_l$. The dependence of $C_v$ on temperature in MX MLs can be observed as shown in Fig. S2. Here, $C_v$ is found to increases as a function of temperature. The PbTe ML is found to exhibit the higher value of $C_v$ than other compounds due to its relatively higher molar mass. At higher temperature, the value of $C_v$ converges to a classical limit of Dulong and Petit, which assumes a constant specific heat for solids.

The lattice thermal conductivity ($\kappa_l$) is an important quantity in the performance of a thermoelectric device, which can be calculated using the following equation [51]:

$$k_l^{\alpha\beta} = \frac{1}{k_B T^2 \Omega N} \sum_\lambda f_0(f_0 + 1)(\hbar w_\lambda)^2 v_\lambda^\alpha F_\lambda^\beta \qquad (3)$$

where $\Omega$ is the volume of the unit cell, $N$ is the number of $q$-points, $f_0$ is the phonon distribution function, $\omega_\lambda$ and $v_\lambda$ are the angular frequency and group velocity of phonon mode $\lambda$, respectively. The calculated $\kappa_l$ at different temperatures are presented in Fig. 4. The figure shows that SnS and SnTe MLs have $\kappa_l$ values of 2.76 (2.94) and 1.47 (1.69) Wm$^{-1}$K$^{-1}$ along armchair (zigzag) direction, respectively, at 300 K. At the same temperature, the values of $\kappa_l$ for PbS and PbTe are 1.16 and 1.41 Wm$^{-1}$K$^{-1}$, respectively. The calculated $\kappa_l$ values of SnS MLs in the present work are in good agreement with the recently reported results 4.4 (4.7) Wm$^{-1}$K$^{-1}$ [35] and 2.95 (3.21) Wm$^{-1}$K$^{-1}$ [41] at 300 K along armchair (zigzag) direction. These $\kappa_l$ values at 300 K are lower for SnTe, PbS and PbTe MLS; and higher for SnS ML in comparison with their bulk forms as shown in Table III. Interestingly, all of these values are very low compared to other 2D materials: graphene (~ 2200 Wm$^{-1}$K$^{-1}$) [67], phosphorene (13.6 Wm$^{-1}$K$^{-1}$) [68], MoS$_2$ (103 Wm$^{-1}$K$^{-1}$), MoSe$_2$ (54 Wm$^{-1}$K$^{-1}$), WS$_2$ (142 Wm$^{-1}$K$^{-1}$) and WSe$_2$ (53 Wm$^{-1}$K$^{-1}$) [20]. The value of $\kappa_l$ decreases with increasing the temperature ($\kappa_l \propto 1/T$ [69]) due to the anharmonic phonon scattering resulting from severe lattice vibration at higher temperature. The lower values of $\kappa_l$ contribute to the higher values of $ZT$, which leads to the better TE performance.

### C. Thermoelectric Properties

The transport properties of a system are evaluated using the distribution function of the electron gas, which tells how the electrons are distributed in the momentum (*k*) space. At equilibrium, the distribution function follows Fermi-Dirac statistics given as [70]:

$$f^{(0)}(\varepsilon) = \frac{1}{exp\left(\frac{\varepsilon - \varepsilon_F}{k_B T}\right) + 1} \qquad (4)$$

where $\varepsilon$, $\varepsilon_F$, $T$ and $k_B$ are the carrier energy, Femi energy, Boltzmann constant and temperature, respectively. There are three possible reasons that account for the change in the electron distribution in *k*-space: the motion of the electrons (diffusion), external forces and scattering processes. The Boltzmann transport theory was developed in order to describe the distribution function in the presence of these effects. By implementing the linearized version of the Boltzmann transport equation (BTE) under relaxation time approximation, the transport distribution function can be calculated as [49]:

$$\sigma(\varepsilon, T) = \int \sum_b v_{b,k}\, v_{b,k} \tau_{b,k} \delta(\varepsilon - \varepsilon_{b,k}) \frac{dk}{8\pi^3} \qquad (5)$$

where $v_{b,k}$, $\varepsilon_{b,k}$ and $\tau_{b,k}$ are the energy, velocity and relaxation time for electron in *b*th band at point *k* in the Brillouin zone, respectively. The moments of transport coefficients can be written as:

$$\mathcal{L}^{(\alpha)}(\mu; T) = q^2 \int \sigma(\varepsilon, T)(\varepsilon - \mu)^\alpha \left(-\frac{\partial f^{(0)}(\varepsilon; \mu, T)}{\partial \varepsilon}\right) d\varepsilon \qquad (6)$$

where $q$ is the electric charge, $f^{(0)}$ is the Fermi-Dirac distribution function, $\mu$ is the chemical potential and $\alpha$ is the index (can have values 0, 1, 2 and so on). Then the electrical conductivity ($\sigma$), electronic thermal conductivity ($\kappa_e$) and the Seebeck coefficient (*S*) can be calculated as:

$$\sigma = \mathcal{L}^{(0)} \qquad (7)$$

$$k_e = \frac{1}{q^2 T}\left[\frac{(\mathcal{L}^{(1)})^2}{\mathcal{L}^{(0)}} - \mathcal{L}^{(2)}\right] \qquad (8)$$

$$S = \frac{1}{qT}\frac{\mathcal{L}^{(1)}}{\mathcal{L}^{(0)}}. \qquad (9)$$

The calculated electrical conductivity ($\sigma$) and electronic thermal conductivity ($\kappa_e$) as a function of deviation of the chemical potential from the Fermi level ($\mu$-$E_F$) at different temperatures 300 K, 500 K and 700 K are shown in Fig. 5. As SnS and SnTe have rectangular structure, the values are shown for the zigzag (dashed line) and armchair (solid line) directions. A significantly high value of $\sigma$ in the order of $10^6$ $\Omega^{-1}$m$^{-1}$ is found near both the conduction band (n-type) and valence band (p-type) (Fig. 5(a-d)). However, the electrical conductivity is lower in MLs than their bulk forms due to the increase in band gaps [11,16,64,71]. In addition, one can notice that $\sigma$ decreases with the increase of temperature, which is associated to the scattering of charge carriers at higher temperature. However, $\kappa_e$ is found to increase with the increase of temperature as more electrons are excited at high temperature (Fig. 5(e-h)). The values of $\kappa_e$ are higher in SnTe and PbTe MLs near the conduction band (for n-type) in comparison with SnS and PbS. The anisotropy in the value of $\sigma$ and $\kappa_e$ for SnS and SnTe MLs can be attributed to the fact that the structures are not cubic ($a_1 \neq a_2$). The calculated $\kappa_e$ values in this work are also found in agreement with the result obtained using Wiedemann-Franz law ($\kappa_e = L\sigma T$) as shown in Fig. S3, where $L$ is the Lorenz number which is $\sim$ 1.5 $\times 10^{-8}$ W$\Omega$K$^{-2}$ for non-degenerate (lightly doped) semiconductors [72].

The Seebeck coefficients ($S$) and power factors ($S^2\sigma$) as a function of $\mu$-$E_F$ at different temperatures are presented in Fig. 6. As the SnX (X = S, Te) MLs have quite similar band distribution along $\Gamma$-X and $\Gamma$-Y directions, the effective mass along both the armchair and zigzag directions should also be similar [36], which in turn leads to less directional dependence of Seebeck coefficient for these compounds as shown in Fig. 6(a) and 6(b). At 300 K, the value of |S| in SnS reaches to the highest peak value ~ 1500 µV/K while SnTe shows the lowest peak value ~ 1200 µV/K among all four compounds. The value of |S| is found to decrease on increasing the temperature, which is attributed to the bipolar conduction [35,73]. The maximum values here are higher than their bulk counterpart [14,16,64]. On the contrary, the power factors ($S^2\sigma$) are found higher at the higher temperatures as affected by the corresponding higher $\sigma$ values (Fig. 6(e-h)). The optimized maximum values of $S^2\sigma$ for p- and n-type of these materials at 700 K are summarized in Table IV. PbS and PbTe MLs have high $S^2\sigma$ values near the valence band (p-type), whereas SnS and SnTe MLs have higher values near the conduction band (n-type). Among all compounds, SnTe shows the maximum $S^2\sigma$ of $77.3\times10^{-4}$ Wm$^{-1}$K$^{-2}$, whereas SnS has the least value of $33\times10^{-4}$ Wm$^{1}$K$^{-2}$. However, higher $S^2\sigma$ values does not necessarily imply the high ZT because of its dependence with $\kappa_e$ and $\kappa_l$ also.

The calculated figure of merit (ZT) under different temperatures as a function of µ-E$_F$ are shown in Fig. 7, where the value of ZT increases with the temperature. The n-type SnS is found to have the maximum ZT value of 1.04 (0.87) at 700 K along armchair (zigzag) direction. However, in the case of SnTe, the armchair and zigzag directions exhibit maximum ZT values of 1.46 and 1.41 for n- and p-type doping, respectively at 700 K. The anisotropy in ZT here is due to the anisotropy in $\sigma$, $\kappa_e$, and $\kappa_l$ values along the armchair and zigzag directions. The highest ZT values of 1.51 and 1.96 are obtained for n-type PbS and p-type PbTe, respectively at 700 K. However, these compounds can be used both as p- or n-type in TE applications as the peaks on ZT along both the conduction and valence bands are significantly high. The maximum values of ZT for the p- and n-type of each of the materials are much higher in comparison with their 3D counterpart as listed in Table IV.

## IV. Conclusion

Structural, electronic, phonon transport and TE properties of 2D SnS, SnTe, PbS and PbTe MLs are investigated using DFT method combined with the Boltzmann transport theory. All compounds show direct band gaps except SnS, which exhibits an indirect band gap. Among all compounds, PbS ML shows the lowest lattice thermal conductivity at all temperatures. The highest *ZT* value of 1.96 was predicted for PbTe ML at 700 K. These materials show good TE properties such as high Seebeck coefficients, high power factors, and low thermal conductivities, which make them promising for efficient TE applications. Further experiments and theoretical studies on these materials with suitable dopings and carrier concentrations could be an important path forward to shed light on TE efficiency enhancement and device applications.

**Acknowledgements**

A. Pandit acknowledges the fruitful discussion with Raad Haleoot. All the calculations were performed through Arkansas High Performance Computing Center at the University of Arkansas.

**Conflict of interest**

The authors confirm that there are no competing interests regarding the materials/work discussed in this manuscript.

**List of Figures:**

**FIG. 1.** Crystal structure of a 2 × 2 × 1 supercell of MX (M = Sn, Pb; X= S, Te) MLs. (a) Top view and (b) Side view of SnS (SnTe); (c) Top view and (d) Side view of PbS (PbTe).

**FIG. 2.** The electronic band structure (SOC included) along $\Gamma$-X-S-Y-$\Gamma$ high-symmetry *k*-points and total density of states (DOS) of (a) SnS, (b) SnTe, (c) PbS and (d) PbTe MLs. The red colored band indicate the valence band maximum and conduction band minimum.

**FIG. 3.** Phonon band structure along $\Gamma$-X-S-Y-$\Gamma$ high-symmetry *k*-points, phonon density of states (PDOS) and group velocity of (a-c) SnS, (d-f) SnTe, (g-i) PbS and (j-l) PbTe MLs. TA1, TA2 and LA represent out of plane transverse, in plane transverse and longitudinal acoustic branches, respectively.

**FIG. 4.** Lattice thermal conductivity ($\kappa_l$) of SnS, SnTe, PbS and PbTe MLs as a function of temperature.

**FIG. 5.** Electrical conductivity ($\sigma$) and electronic thermal conductivity ($\kappa_e$) as a function of deviation of the chemical potential from the Fermi level ($\mu$-$E_F$) at different temperatures 300 K, 500 K and 700 K for (a and e) SnS, (b and f) SnTe, (c and g) PbS and (d and h) PbTe MLs. Solid lines represent the values in armchair directions and the dashed lines represent the values in zigzag directions (for SnS and SnTe).

**FIG. 6**. Seebeck coefficient (*S*) and power factor ($S^2\sigma$) as a function of deviation of the chemical potential from the Fermi level ($\mu$-$E_F$) at different temperatures 300 K, 500 K and 700 K for (a and e) SnS, (b and f) SnTe, (c and g) PbS and (d and h) PbTe MLs. Solid lines represent the values in armchair directions and the dashed lines represent the values in zigzag directions (for SnS and SnTe).

**FIG. 7.** Calculated TE figure of merit (*ZT*) as a function of deviation of the chemical potential from the Fermi level ($\mu$-$E_F$) at different temperatures 300 K, 500 K and 700 K for (a) SnS, (b) SnTe, (c) PbS and (d) PbTe MLs. Solid lines represent the values in armchair directions and the dashed lines represent the values in zigzag directions (for SnS and SnTe).

**List of Tables:**

**TABLE I.** Optimized lattice constants of bulk and layered MX (M = Sn, Pb; X= S, Te) compounds along with previous results.

**TABLE II.** The electronic band gap of MX (M = Sn, Pb; X= S, Te) compounds in their bulk and layered forms along with previous results. Values in parenthesis indicate the band gaps without SOC.

**TABLE III**. Comparison of lattice thermal conductivities ($\kappa_l$) of MX (M = Sn, Pb; X= S, Te) MLs with their bulk forms at 300 K.

**TABLE IV.** Maximum values of power factor ($S^2\sigma$) and the TE figure of merit (*ZT*) for *p*- and *n*-type of SnS, SnTe, PbS and PbTe MLs along with their bulk forms.

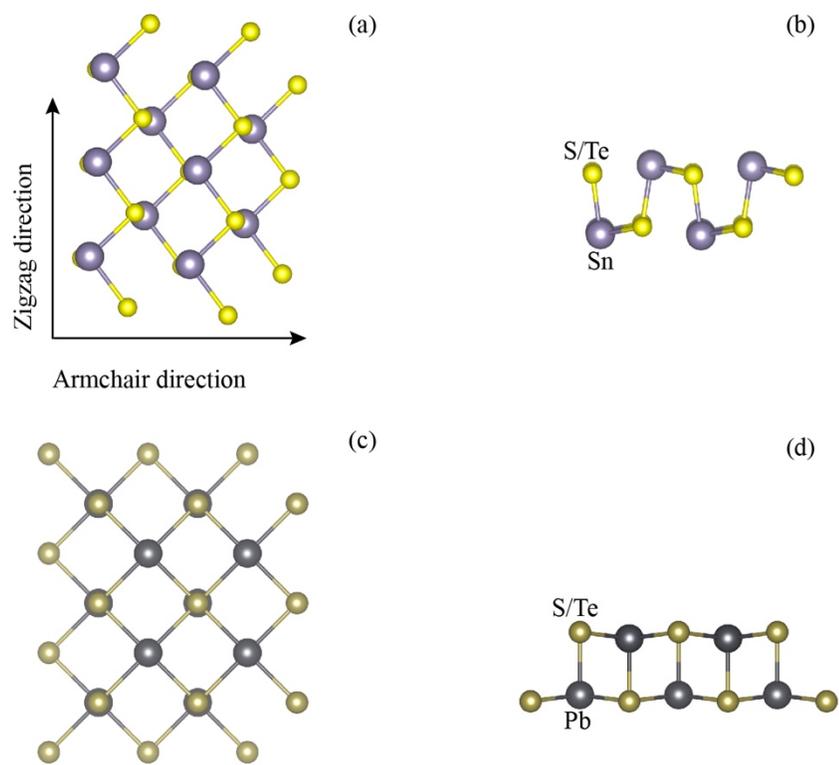

**FIG. 1**

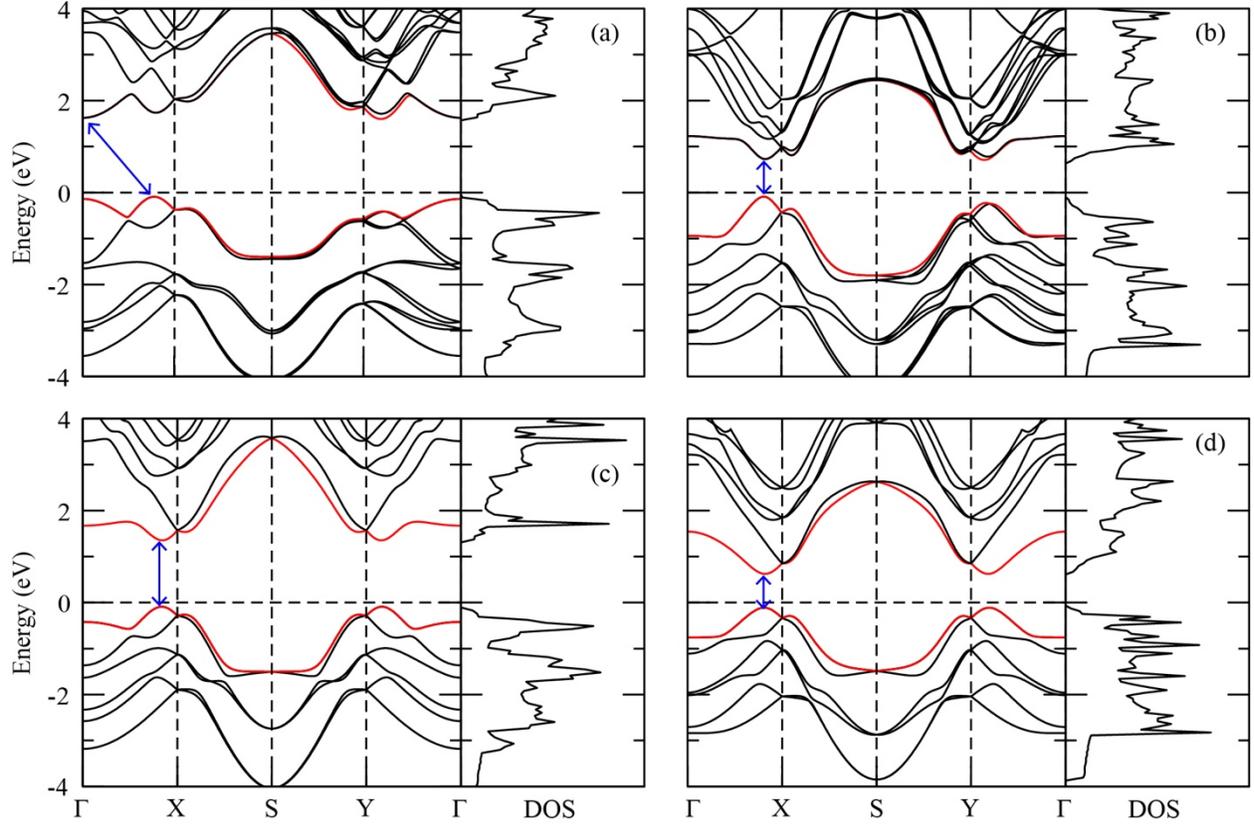

**FIG. 2**

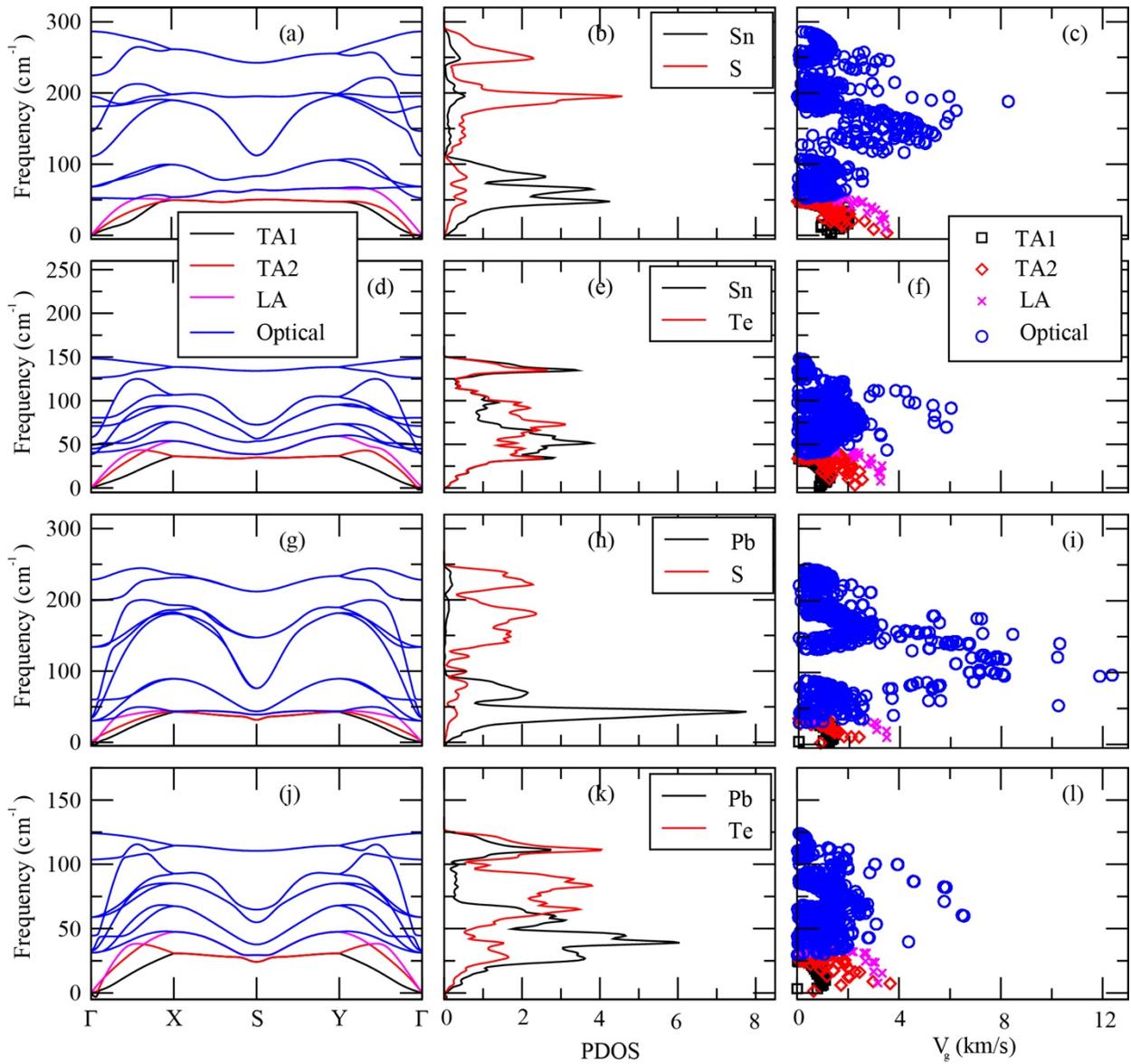

**FIG. 3**

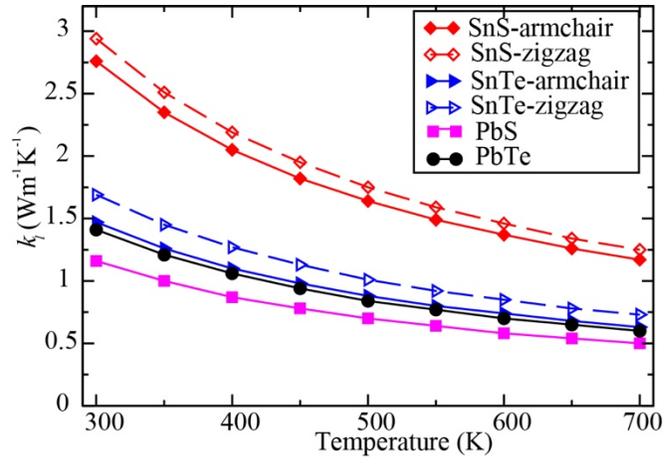

**FIG. 4**

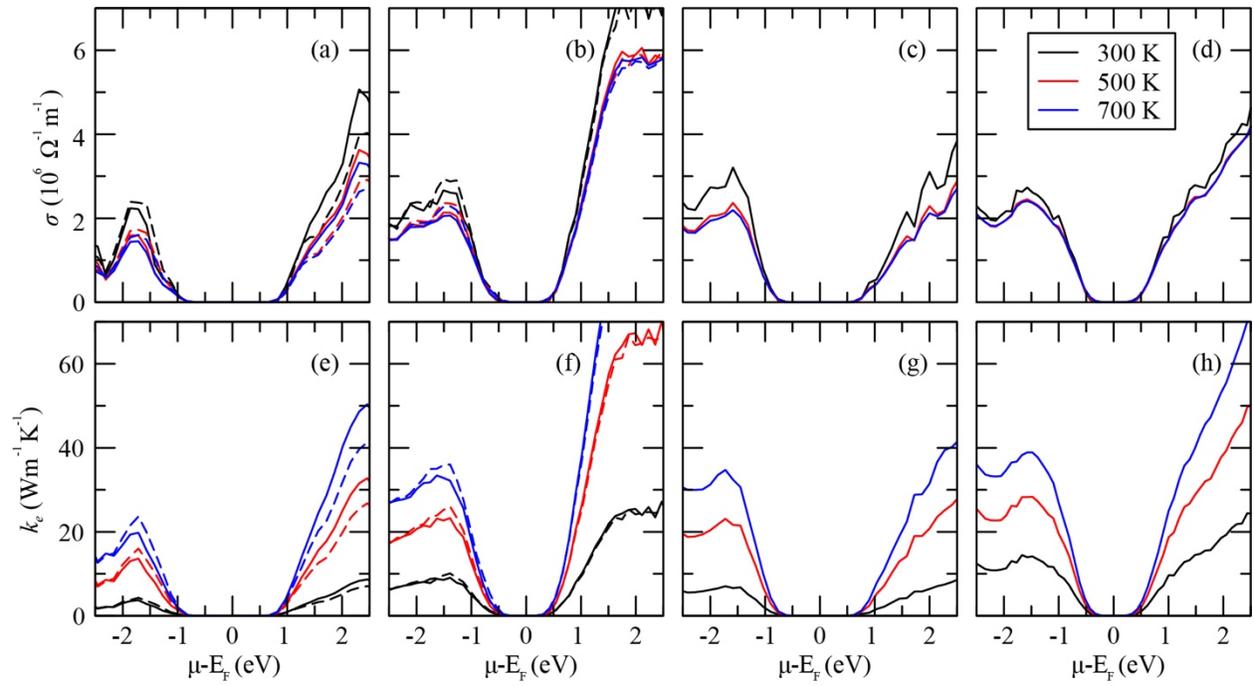

**FIG. 5**

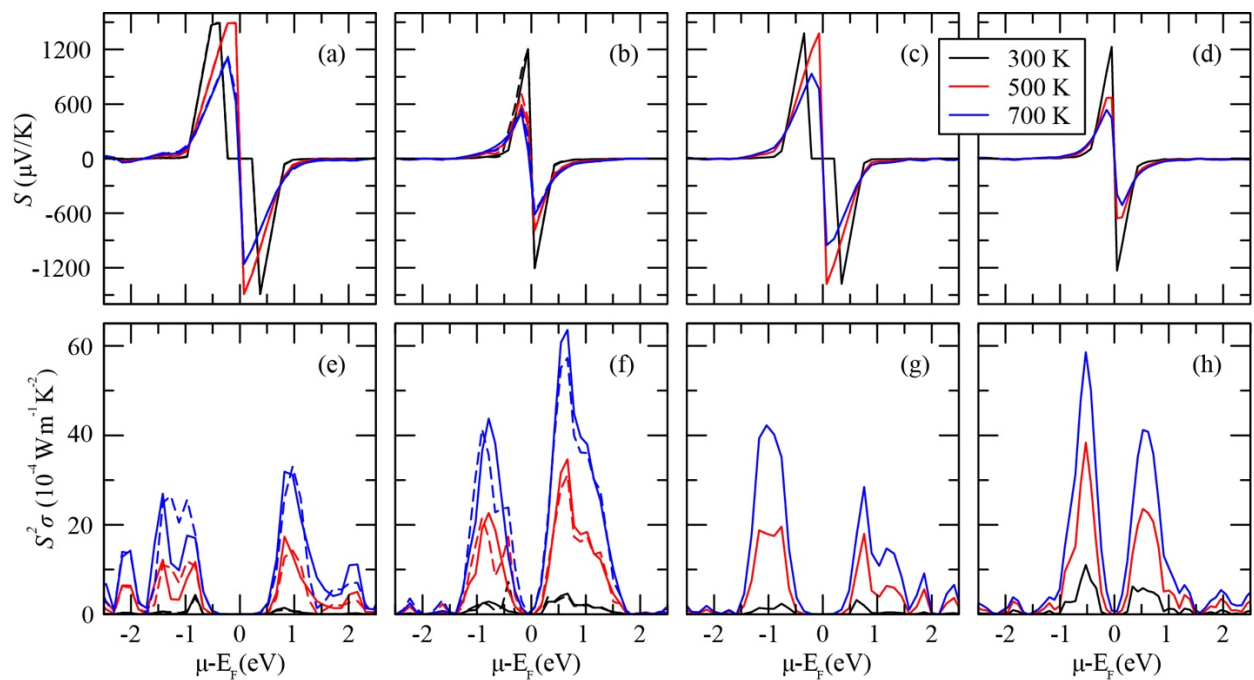

**FIG. 6**

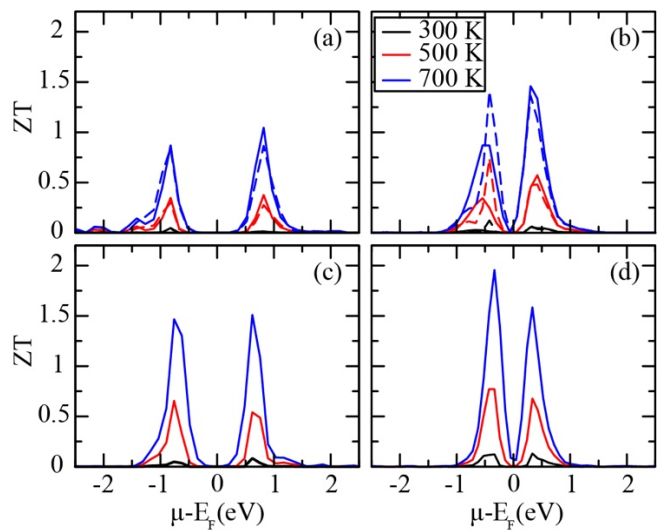

**FIG. 7**

TABLE I

| Compounds | Lattice constant (Å) | | | |
|---|---|---|---|---|
| | Bulk | | Monolayer | |
| | This work | Previous results | This work | Previous results |
| SnS | 11.41, 4.02, 4.43 | 11.4, 4.04, 4.35 [16] | 4.33, 4.08 | 4.31, 4.07 [35] <br><br> 4.24, 4.07 [53] |
| SnTe | 6.40 | 6.41 [54] | 4.66, 4.58 | 4.73, 4.57 [39] <br><br> 4.54, 4.58 [55] |
| PbS | 5.99 | 5.94 [56] <br><br> 5.99 [57] | 4.25 | 4.23 [58] |
| PbTe | 6.56 | 6.56 [54,57] | 4.67 | 4.65 [59] <br><br> 4.63 [58] |

TABLE II

| Compounds | Band gap (eV) | | | |
|---|---|---|---|---|
| | Bulk | | Monolayer | |
| | This work | Previous results | This work | Previous results |
| SnS | 1.13 | 0.91 [16] | 1.69 (1.62) | 1.96 [63] <br> 1.52 [58] |
| SnTe | 0.106 | 0.097 [64] | 0.79 (0.86) | 0.7 [55] <br> 0.72 [58] |
| PbS | 0.113 | 0.074 [57] | 1.44 (1.73) | 1.66 [58] |
| PbTe | 0.116 | 0.091 [57] | 0.73 (1.31) | 1.26 [59] <br> 1.20 [58] |

**TABLE III**

| Compounds | $\kappa_l$ (Wm$^{-1}$K$^{-1}$) | | |
|---|---|---|---|
| | Monolayer | | Bulk |
| | Armchair | Zigzag | |
| SnS | 2.76 | 2.94 | 0.45 [16] |
| SnTe | 1.47 | 1.69 | 2 [64] |
| PbS | 1.16 | | 2.5 [14] |
| PbTe | 1.41 | | 2 [6] |

**TABLE IV**

| Compounds | Monolayer | | | Bulk |
|---|---|---|---|---|
| | Type | $S^2\sigma$ (10$^{-4}$ Wm$^{-1}$K$^{-2}$) | ZT | ZT |
| SnS | p | 33.38 | 0.87 | ~ 1 (Only $\kappa_e$ used) [16] |
| | n | 33 | 1.05 | |
| SnTe | p | 43.76 | 1.41 | 0.3 [64] |
| | n | 77.3 | 1.46 | |
| PbS | p | 42.24 | 1.47 | 0.96 [14] |
| | n | 28.5 | 1.51 | |
| PbTe | p | 58.61 | 1.96 | 0.77 [74] |
| | n | 41.22 | 1.59 | |